# Angular-Dependent Dynamic Response and Magnetization Reversal in Fibonacci-Distorted, Kagome Artificial Spin Ice

Ali Frotanpour, Justin Woods, Barry Farmer, Amrit P. Kaphle and Lance E. DeLong

Department of Physics and Astronomy, University of Kentucky, 505 Rose Street, Lexington, Kentucky 40506-0055, USA

## Abstract:

We have measured the angular dependence of ferromagnetic resonance (FMR) spectra for Fibonacci-distorted, Kagome artificial spin ice (ASI). The number of strong modes in the FMR spectra depend on the orientation of the applied DC magnetic field. In addition, discontinuities observed in the FMR field-frequency dispersion curves also depend on DC field orientation, and signal a multi-step DC magnetization reversal, which is caused by the reduced energy degeneracy of Fibonacci-distorted vertices. The results suggest the orientation of applied magnetic field and severity of Fibonacci distortion constitute control variables for FMR modes and multi-step reversal in future magnonic devices and magnetic switching systems.

## Introduction:

Artificial spin ice (ASI) is a geometrically confined, 2D magnetic nanostructure comprised of a lattice of sub-micron, ferromagnetic thin-film segments [1,2]. These segments have strong shape anisotropy (length l >> width w >> thickness t), which constrains them to behave as a single-domain, Ising-like spin. The forces between nearest-neighbor segments are dominated by dipolar interactions. However, exchange interactions play an important role in forming the magnetization textures near vertices in connected ASI [1-4]. The competition between magnetostatic interactions among nearest-neighbor segments leads to degenerate ground states that follow the spin ice rule (SIR) similar to the original rule observed for atomic spins in the tetrahedral sublattice of pyrochlore spin ice [5]. In the last decade, researchers have studied ASI in variety of lattice types such as honeycomb (Kagome ASI) [6-18], square [19-22] and artificial quasicrystals [23-29].

Ferromagnetic resonance (FMR) spectroscopy is a powerful technique for characterizing the dynamic magnetic response in ASI. FMR spectroscopy provides detailed information about magnetization texture and shape anisotropy of ASI. For example, magnetic shape anisotropy can be estimated from the slope of the frequency-field graph, df/dH (i.e., the steeper the slope, the stronger the shape anisotropy) [14]. We can also find the reversal fields for sublattices of segments by observing two or more discontinuities in f(H) or df/dH, or mode softening in f(H) [29].

In recent years, researchers characterized the FMR modes of ASI in different lattice types [11-18, 21, 22, 26-29]. Kagome ASI is particularly important since it possesses strongly-frustrated, threefold vertices that resemble the threefold tetrahedral bonds of pyrochlore compounds [5]. Also, the FMR spectrum of Kagome ASI strongly depends on the orientation of the magnetic field such that the number of strong modes in the field-frequency dispersion curves (i.e., degeneracy) can be controlled by orienting of the applied DC magnetic field [11-13]. From a technology perspective, the spin wave spectrum of Kagome ASI is very rich, which is desirable for applications in magnonic devices [30-34]. Moreover, Kagome ASI exhibits a two-step reversal process that depends on the angle of applied DC magnetic field [8, 12], which also has potential applications in magnetic switching systems and spintronics devices [13, 26, 35].

Recently, we introduced a distorted Kagome ASI created by applying an algorithmic Fibonacci sequence to modulate the honeycomb lattice parameters [29]. We showed that the distortion changes the reversal behavior; and the FMR modes can be tuned by the severity of distortion (distortion ratio, $r$), which can only be varied as part of the sample fabrication process.

In this paper, we employ broadband FMR spectroscopy with VNA detection, and micromagnetic simulations to show that the FMR modes and magnetization reversal process in the Fibonacci-distorted Kagome ASI can be controlled by a convenient, adjustable parameter --- i.e., the orientation of the applied DC magnetic field with respect to the ASI lattice. The dependence of the number of observed strong FMR modes, and the multi-step reversal on applied DC field orientation, offer an improved paradigm for magnonic devices [30-34].

**Methods**:

The geometry and algorithm for fabricating the Fibonacci-distorted Kagome ASI are described in [29]. We define a ratio $r$ = L/S of the lengths of long to short film segments, such that $r$ = 1.0 represents an undistorted, periodic Kagome ASI, and $r > 1$ quantifies the severity of the loss of periodic lattice symmetry in distorted samples. **Figures 1 (a)** and **(b)** illustrate distortion ratios $r$ = 1.3 and 1.62, respectively. Samples were patterned using electron beam lithography on a quartz ($SiO_2$) substrate, followed by deposition of 25 nm of Permalloy in an electron beam evaporator system, and a lift-off procedure. An SEM image of a sample fabricated with $r$ = 1.62 is shown in **Figure 1 (c)**. The patterned sample consisted of a hexagonal write field of approximately 43 × 43 microns, repeated in a grid of 4 × 80 hexagons with a lattice spacing of 100 microns. The undistorted hexagonal pattern has a segment length l = 500 nm and width W = 140 nm.

We distinguish between subsets of Ising segments based on their direction with respect to the easy *x*-axis (see **Figure 1 (c)**), and label them with capital letters A through E, as shown in **Figure 1 (c)**. Segments A, B (D), C (E) for $r$ = 1.62 correspond to $\phi$ = ~0°,+47° (-47°) and +60° (-60°), respectively.

Broadband FMR spectroscopy was performed by placing the sample in a "flip-chip" on a microstripline and collecting data for the transmission coefficient ($S_{12}$) measured by a vector network analyzer (VNA) in the frequency range of 4 GHz to 14 GHz for a fixed DC magnetic field applied at angle $\phi$, as shown in **Figure 1 (c)**. A detailed description of the FMR setup can be found in [29]. For each sample, the magnetic field was swept from +900 Oe to -900 Oe for $\phi$ = 15°, 30° and 45°. This range of applied DC fields includes the Ising saturated state and magnetization reversals. We subtracted the $S_{12}$ data taken at +3000 Oe from those taken at each applied DC field to remove any background signal.

We used the Object Oriented Micromagnetic Framework (OOMMF) to simulate the equilibrium magnetization texture and FMR spectrum. We used a pixel size of 10 × 10 × 25 $nm^3$, and a Permalloy saturation magnetization $M_s$ = 800 kA/m and exchange stiffness $A = 1.3 \times 10^{-12}$ J/m [35] . We recorded the magnetization vectors for each pixel generated by OOMMF and used them as input to simulate the FMR signal. We used a 10-Oe magnetic field pulse of 20 ps duration applied perpendicular to the plane of the film to excite FMR. Magnetization vectors were recorded every 20 ps over a 20-ns time interval. We then applied a fast-Fourier-transform

(FFT) of the time-dependent magnetization for each pixel. Finally, the FFT of all pixels were averaged to simulate the observed FMR absorption spectrum.

## Results:

FMR modes in ASI are typically described in a frequency-field graph that includes the Ising saturation and reversal fields. In the Ising saturation regime, the slopes of the frequency-field graphs (df/dH) depend on the slowly saturation magnetization texture [11-18]. Alternatively, modes in the reversal regime indicate more rapid changes in the magnetization configuration. We observe that FMR modes in the Ising and reversal regimes are sensitive to the orientation of the applied field and the distortion ratio, $r$. In what follows, we first discuss the dependence of the FMR modes on the applied field angle in the Ising regime. Then we explain changes in the FMR modes in the reversal regime with the help of micromagnetic simulations.

### a. A tunable FMR spectrum in Ising regime

**Figure 2** and **Figure 3** show FMR experiment and simulation results for $r = 1.3$ and $r = 1.62$, respectively, for applied field angles $\phi = 15°$, $30°$ and $45°$. The number of modes, their frequencies and df/dH are all affected by the change in the distortion ratio, $r$ and the angle $\phi$. For $r = 1.3$ in the Ising saturation regime (close to 900 Oe), three modes are distinguished at $\phi = 15°$ for frequencies above 8 GHz, whereas only two modes are discerned at 30° and 45°. For $r = 1.62$ and frequencies above 8 GHz in the Ising saturation regime, we observed five, three and four modes for $\phi = 15°$, 30° and 45°, respectively. These data demonstrate that we can adjust the number of modes from two up to five by simply changing the applied field orientation. We conclude that the application of a Fibonacci distortion to the Kagome ASI generates a lattice with a highly tunable frequency spectrum.

Our simulations indicate modes active within the vertex regions are observed below 8 GHz, and can be categorized as localized domain wall modes (LDW), or vertex center modes (VCM) [29]. We identified LDW modes in experiment and simulations for all ratios and angles studied; however, we observed the VCM in experiment and simulations only for $\phi = 45°$ (see **Figure 2** and **Figure 3**), as we will discuss in more detail, below.

We can locate FMR modes using simulated maps of the FMR power absorption (mode profile). Characterization of the FMR modes in ASI is fundamentally important since it reveals details of the magnetization texture and internal magnetic field, and it is instructive for the design of magnonic devices. **Figures 4 (a)-(e)** show that the mode profiles (labeled A-E in **Figure 3 (a)**) for $r = 1.62$ with applied field at 900 Oe and $\phi = 15°$ can be assigned to resonances of Segments A-E (Labeled in **Figure 1 (c)**), respectively.

We observed that the VCM exists for the applied field oriented at $\phi = 45°$, and is absent when $\phi = 15°$ and $\phi = 30°$. We have previously shown [4] that the existence of this mode depends on two conditions: (1) At least one segment's easy axis must be in the direction of the applied DC magnetic field. (2) The DC applied field orientation and average vertex magnetization direction must be roughly aligned. We now observe that condition (1) is not necessary. The VCM mode profile and vertex magnetization texture shown in **Figure 5** suggest that the VCM exists if the orientation of the magnetization in the center of the vertex region is near to 45°, which is the same as the orientation of the applied field. (Note the VCM shown in **Figure 5** exists when none of the near-neighbor segment's easy axes are aligned with the applied DC field.) Taken together, these results indicate that the existence of VCM is only sensitive to the ***relative orientation of DC applied field and vertex magnetization***: Note that the VCM profile has approximate local mirror symmetry with respect to the applied DC field at 45°, and aligns with the local average magnetization texture; whereas, the distorted Kagome lattice has only global mirror symmetry with respect to the 0° axis, as does the VCM profile when the applied DC field is aligned at 0° (see [29], Figure 8). These observations reflect the "local" symmetry of the VCM when the applied DC field direction violates the global lattice mirror symmetry.

*b. Multi-step magnetization reversal process*

Understanding the magnetization reversal process in ASI is critical to the development of magnetic switching applications [8, 11], and elucidates the physics of the interactions between the Ising segments. Kagome ASI exhibits a two-step reversal process when the applied field is oriented at certain angles. We observe that the reversal in the Fibonacci-distorted, Kagome ASI can occur in multiple steps, and originates in vertices with symmetry reduced by the distortion. The FMR spectra for $r = 1.62$ and $\phi = 15°$ (**Figure 3(a)**) reveal the existence of Modes M and N during reversal in a range of fields spanning -200 Oe to -600 Oe, while Mode E reappears after reversal at a DC field near -650 Oe, in agreement with simulations. The multi-step reversal process is determined primarily by abrupt changes in the magnetization textures of subgroups of segments that simultaneously reverse at particular "critical fields", as shown in **Figure 6**.

Before we discuss the details of reversal, it is important to note that there are 8 different vertex types in the Fibonacci-distorted Kagome ASI (see Supplementary Information (SI), **Figures S1 and S2**, and the energies for each vertex for all magnetization configurations are shown in SI, **Table S1**). The Fibonacci distortion breaks the sixfold symmetry of the array of threefold vertices in the non-distorted Kagome ASI. As a result, four vertices with fully broken symmetry (see SI, **Figure S2**) have four non-degenerate energy levels, but remain consistent with SIR ("one-in (out)-two-out (in)").

The remnant state after Ising saturation with the applied DC field at $\phi = 15°$ is shown in **Figure 6 (a)**. As can be seen, not all vertices have relaxed into their ground-state texture, which indicates that nearby segments will have to reverse at relatively low applied fields to trigger the reversal of high-energy vertices that persist on the boundaries of the Kagome write fields, as well as those in the second excited energy state (i.e., vertex $V_a$ at -390 Oe, as shown in **Figure 6 (a)**). Alternatively, the reversals of segments that neighbor vertices $V_h$ and $V_e$ (labeled in **Figure 6 (b)**) are blocked, because they must enter a higher energy state after reversal of Segments $S_h$ and $S_e$.

The next reversal event occurs at -480 Oe (**Figure 6 (c)**), and involves segments that satisfy two conditions: (1) The angle of their Ising axis with respect to the *x*-axis is close to $\phi = 15°$. (2) The angle between the segments (labeled $\theta_1$ in **Figure 6 (c)**), is larger than the angle between non-reversed segments (labeled $\theta_2$ in **Figure 6(c)**).

The next series of segments reverse at -580 Oe (shown in **Figure 6 (d)**). These segments make an angle $\theta_2$ (see **Figure 6 (c)**) between them. Finally, the segments that make angles near 90° with respect to the applied field (segments labeled S in **Figure 6 (d)**) reverse at -620 Oe.

**Conclusion:**

We have measured the angular-dependent FMR spectrum for Fibonacci-distorted Kagome ASI using broadband FMR spectroscopy and micromagentic simulations. We have observed novel alterations of the FMR spectra in response to small changes of the DC field orientation and the degree of Fibonacci distortion. We found that adjusting the angle of the applied DC magnetic field can alter the number of strong bulk modes (residing inside the body of a film segment) over a range of two to five. We showed that the magnetization reversal process in the Fibonacci-distorted Kagome ASI follows a multi-step behavior for a narrow range (approximately ±5°) of applied DC field directions near $\phi = 15°$. The variable number of strong FMR modes and the complicated reversal process are direct consequences of the reduced symmetry defined by the Fibonacci algorithm, and pose potential applications for magnonic filters and control of magnetic switching in devices [30-34].

**Acknowledgments:**

Research at the University of Kentucky was supported by the US NSF Grant No. DMR-1506979, the UK Center for Advanced Materials, the UK Center for Computational Sciences, and the UK Center for Nanoscale Science and Engineering. Research at the Argonne National Laboratory, a US Department of Energy Office of Science User Facility, was supported under Contract No. DE-AC02-06CH11357

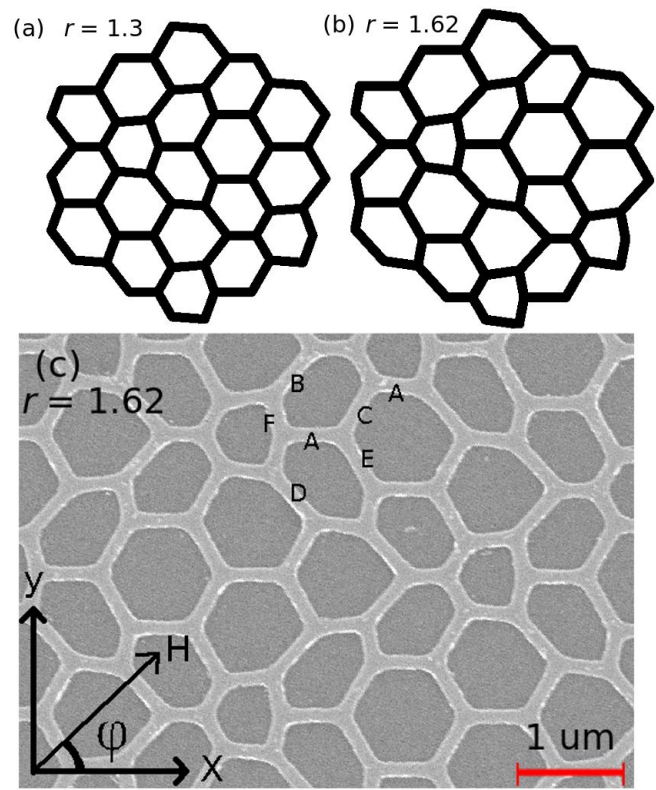

**Figure 1. (a)** and **(b).** Fibonacci-distorted Kagome ASI with $r = 1.3$ and $r = 1.62$, respectively. **(c)** Shows a sample with $r = 1.62$. $\phi$ is the angle between the applied DC magnetic field and the *x*-axis. Segments are labeled according to the angle between their major axis and *x*-axis.

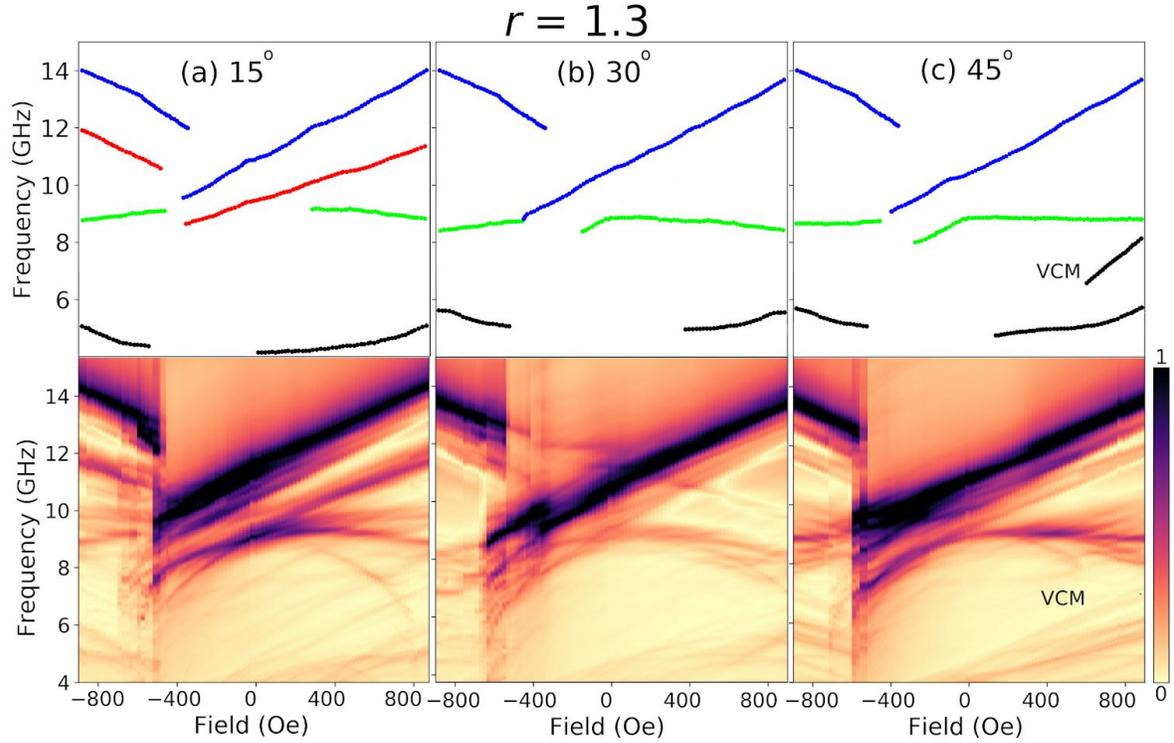

**Figure 2.** (a)-(c) Experimental (top panels) and simulated (bottom panels) FMR spectra for Fibonacci-distorted Kagome ASI with $r$ = 1.3 for $\phi$ = 15°, 30° and 45°, respectively. The number of modes and their slope change significantly with the applied DC field orientation. The Fibonacci distortion reduces the symmetry of the Kagome ASI, which leads to different FMR spectra for $\phi$ = 15° and 45°. The VCM is only found for $\phi$ = 45°. The lowest frequency mode is the localized domain wall mode (LDW).

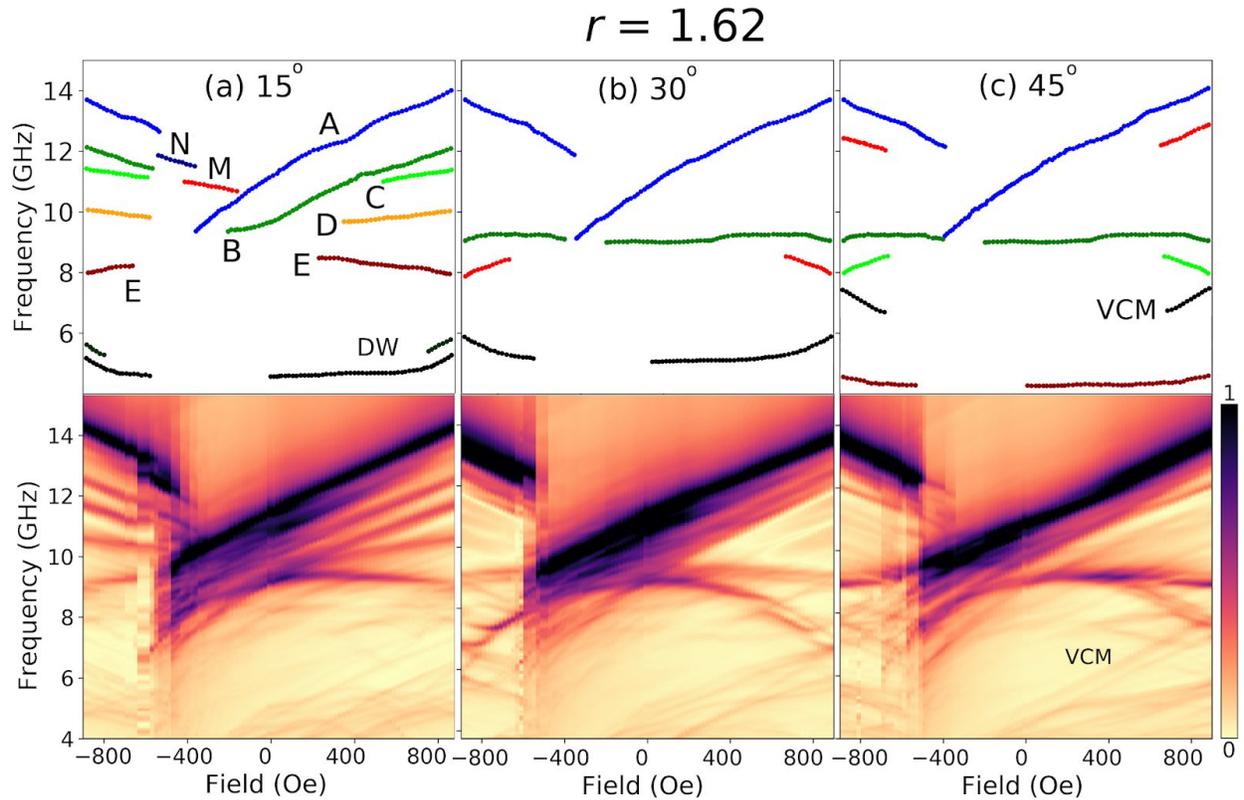

**Figure 3. (a)-(c)** Experimental (top panels) and simulated (bottom panels) FMR spectra for Fibonacci-distorted Kagome ASI with $r$ = 1.62 for $\phi$ = 15º, 30º and 45º, respectively. The number of modes and their slope change significantly with the applied field orientation; specifically for $\phi$ = 15º we can distinguish five modes near 900 Oe (at Ising saturation). The VCM is only found for $\phi$ = 45º. The lowest frequency mode is the localized domain wall mode (LDW). We observe multi-step reversal behaviour for $\phi$ = 15º, where Modes N and M appear, and Mode E reappears at higher fields.

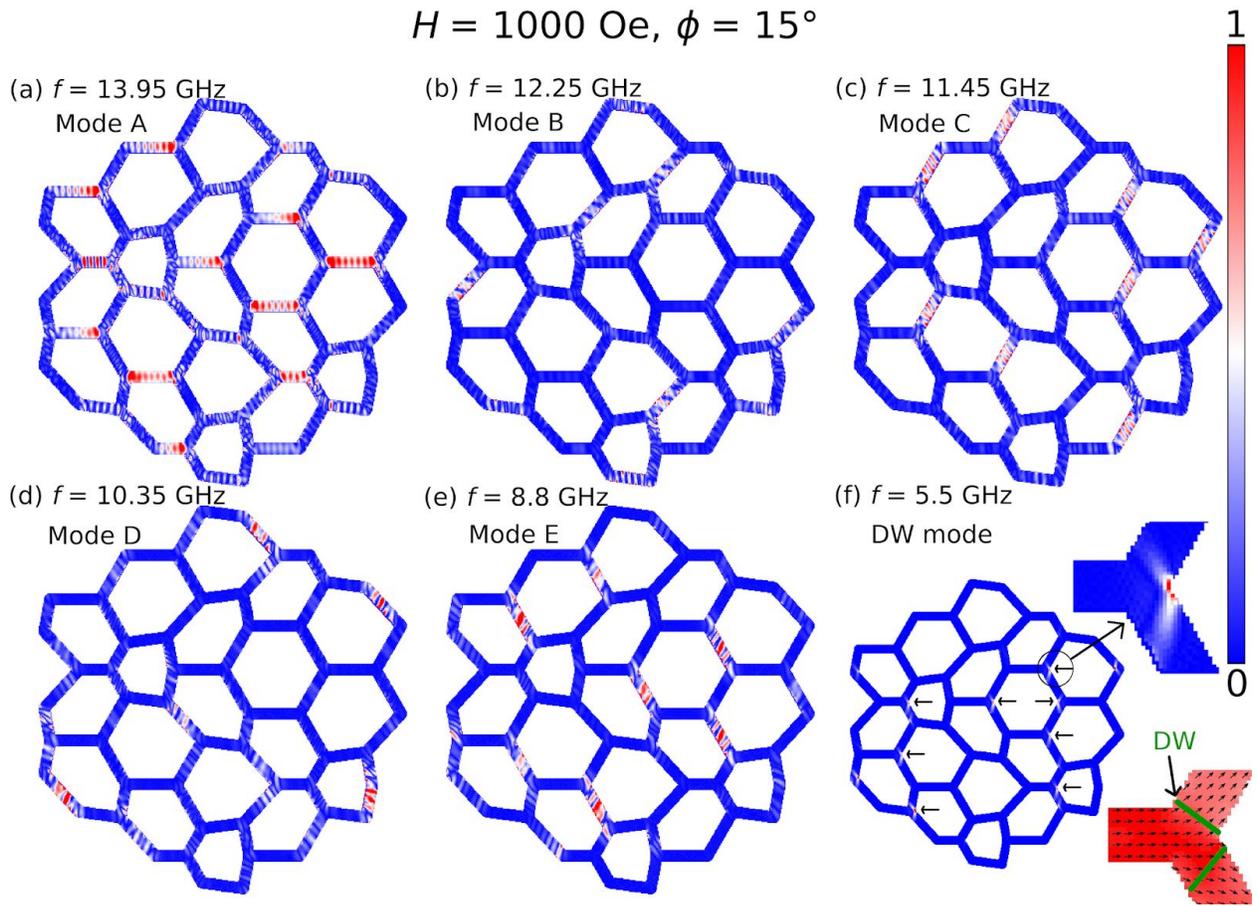

**Figure 4. (a)-(f)** FMR Mode profiles for Modes A, B, C, D, E, and DW, respectively. Power absorption maps are plotted at the frequencies where the simulated FMR spectrum peaks for a distortion ratio $r = 1.62$, applied DC magnetic field $H = 1000$ Oe and $\phi = 15°$. In **(f)**, the DW is shown and confirms that the LDW mode is sensitive to the DW location. The colorbar (shown on the right side) depicts red as the maximum absorption intensity and blue as zero absorption.

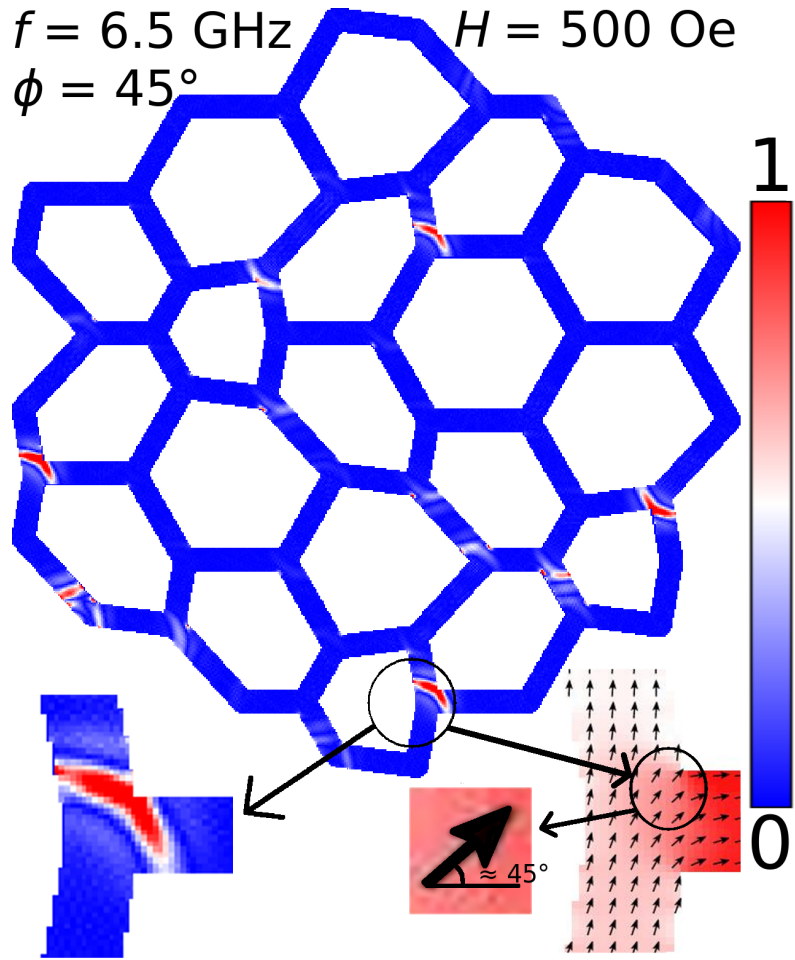

**Figure 5.** VCM profile is plotted for 6.5 GHz, for distortion ratio of $r = 1.62$ and applied DC magnetic field of $H = 500$ Oe ($\phi = 45°$). This mode is absent at other applied field orientations. A magnified VCM profile can be seen on the bottom left, and demonstrates local mirror symmetry of this profile with respect to the applied field direction (45°). Note that the magnetization texture in the vertex center is also symmetrically aligned with applied DC field direction (as shown by the large black arrow in the bottom right). The color bar at the right side shows red as the maximum absorption intensity and blue as zero absorption.

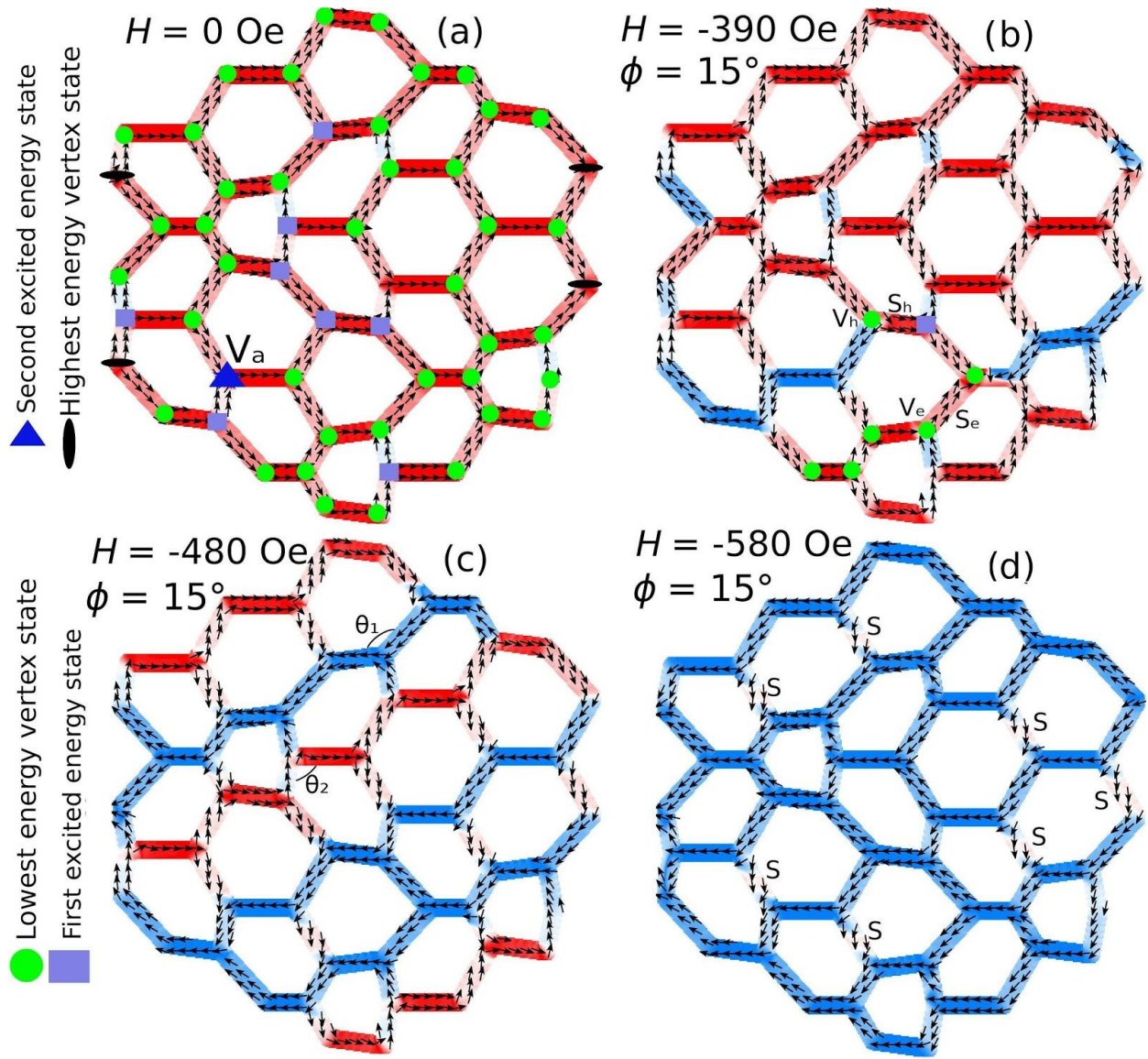

**Figure 6. (a)** Remnant state at zero field after Ising saturation at applied DC field angle $\phi = 15^0$. Ground state vertices are shown as green circles, and vertices in the highest energy state are shown as black ellipses. First excited states are shown as purple squares, and the second excited states are shown as blue triangles. Vertices with fully broken symmetry have four non-degenerate energy levels. One vertex in the second excited state is labeled as $V_a$. **(b)-(d)** Magnetization textures just after each reversal. **(b)** shows the magnetization texture at -390 Oe. The highest energy vertices (e.g., vertex $V_a$ shown in (a)) trigger reversal of segments shown in (b). The segment $S_h$ and $S_e$ block further reversal since vertices $V_h$ and $V_e$ must switch to a higher energy state if those segments

reverse. In **(c)**, segments that make angle ~$\theta_1$ reverse at -480 Oe. In **(d),** segments that make an angle of ~$\theta_2$ reverse at -580 Oe. Note $\theta_1 > \theta_2$. Segments labeled S in **(d)** reverse at -620 Oe, since they make the largest angle between their magnetization and applied magnetic field, as compared to other segments.